\documentclass[twocolumn, aps, prl, 10pt]{revtex4}

\usepackage{siunitx}
\usepackage{graphicx}
\usepackage{amsmath,amsfonts}
\usepackage{lipsum}
\usepackage{physics}

\usepackage{cleveref}
\usepackage{float}
\usepackage{xfrac}
\usepackage[usenames, dvipsnames]{color}
\usepackage{cancel}
\usepackage{xcolor}
\usepackage{array}
\usepackage{comment}
\usepackage{tabu}
\usepackage{rotating}
\usepackage{appendix}

\newcolumntype{L}[1]{>{\raggedright\let\newline\\\arraybackslash\hspace{0pt}}m{#1}}
\newcolumntype{C}[1]{>{\centering\let\newline\\\arraybackslash\hspace{0pt}}m{#1}}
\newcolumntype{R}[1]{>{\raggedleft\let\newline\\\arraybackslash\hspace{0pt}}m{#1}}

\begin{document}

\title{Quantum-limited determination of refractive index difference by means of entanglement}

\author{M. Reisner$^{1}$, F. Mazeas$^{1}$, R. Dauliat$^{2}$, B. Leconte$^{2}$, D. Aktas$^{1,\dagger}$, R. Cannon$^{1}$, P. Roy$^{2}$, R. Jamier$^{2}$, G. Sauder$^{1}$, F. Kaiser$^{1,\dagger\dagger}$, S. Tanzilli$^{1}$, and L. Labont\'e$^{1}$}

\affiliation{$^{1}$Universit\'e C\^ote d'Azur, CNRS, Institut de Physique de Nice, 06108 Nice Cedex 2, France.}
\affiliation{$^{2}$Universit\'e de Limoges, XLIM, UMR 7252, Limoges, France.}
\affiliation{$^{\dagger}$Now at Quantum Engineering Technology Labs, University of Bristol, UK}
\affiliation{$^{\dagger\dagger}$Now at Center for Integrated Quantum Science and Technology, University of Stuttgart, Germany.}

\date{\today}

\begin{abstract}{Shaping single-mode operation in high-power fibres requires a precise knowledge of the gain-medium optical properties. This requires accurate measurements of the refractive index differences ($\Delta n$) between the core and the cladding of the fiber. We exploit a quantum optical method based on low-coherence Hong-Ou-Mandel interferometry to perform practical measurements of the refractive index difference using broadband energy-time entangled photons. The precision enhancement reached with this method is benchmarked with a classical method based on single photon interferometry. We show in classical regime an improvement by an order of magnitude of the precision compared to already reported classical methods. Strikingly, in the quantum regime, we demonstrate an extra factor of 4 on the accuracy enhancement, exhibiting a state-of-the-art $\Delta n$ precision of $\num{6e-7}$. This work sets the quantum photonics metrology as a powerful characterization tool that should enable a faster and reliable design of materials dedicated to light amplification.} 
\end{abstract}

\maketitle
\section{Introduction}
Fiber light sources are among key-growth technologies in the field of photonics owing to their outstanding performance in terms of high average power, excellent beam quality, single- and multi-pass gain, and agility \cite{Richardson:10}. They have revolutionized existing scientific and industrial applications in the biomedical field, and industrial materials processing for example, as well initiate new ones, as metrology and imaging~\cite{noauthor_extending_2012, noauthor_fibre_2013}. Fiber laser development relies on a  complementary approach between tailored waveguide design and low-loss optical materials synthesis for enabling high-power propagation. Much effort has been devoted to waveguide engineering, leading to speciality fibre architectures such as microstructured very large-area-mode fibers (VLMA)~\cite{kalli_large_2015, Dauliat16}. Optical materials have also received a great attention through dedicated engineering work~\cite{dragic_materials_2018, schuster_material_2014}. However, despite the progress made over the last decade, an experimental method allowing precise characterization of optical material properties is still missing. A striking example is that of VLMA fibers. The cornerstone of their fabrication lies in the precise knowledge of the refractive index difference $\Delta n$ between the two different materials composing the core and the cladding of the waveguide, which has to be lower than $10^{-5}$ to ensure single-mode operation within a large core fiber~\cite{Dauliat16}. The associated precision should be at least one order of magnitude lower, \textit{i.e} $\sim 10^{-6}$. Unfortunately, state-of-art precision achievements based on optical coherence tomography (OCT) are limited to $\num{e-4}$~\cite{Tan:16,singh2006, Yablon:13}, mainly due to chromatic dispersion.\\
In this paper, we introduce an experimental method based on quantum OCT allowing measurements of $\Delta n$ with a precision down to $\num{6e-7}$, corresponding to a 4-fold enhancement with respect to classical methods. This consists in exploiting an Hong-Ou-Mandel (HOM)-type interferometer fed with low-coherence energy-time entangled photons ~\cite{HOM}. In comparison to single-photon based experiments, exploiting quantum biphoton states exhibits two main advantages~\cite{abouraddy_quantum-optical_2002}: (i) the instrument’s resolution is not affected by even-order dispersion in the sample thanks to dispersion cancellation resulting from the energy correlation, and (ii) an augmented robustness to the losses of the sample under test (SUT). In addition to the increased precision, this approach is independent on the SUT, leading to universal and versatile optical property measurements~\cite{kaiser_quantum_2018}. \\
HOM-interferometry stands as a fundamental concept in quantum optics~\cite{HOM} and is of particular relevance for the measurement of indistinguishable photons~\cite{mcmillan_two-photon_2013}, that lies at the heart of quantum teleportation and entanglement swapping~\cite{halder_entangling_2007, dauria_universal_2020}. Furthermore, the HOM effect has been exploited for generating path-entangled two-photon N00N state~\cite{Hua:21}, a class of states widely used in enhanced phase-sensing based quantum-metrology. This includes microscopy~\cite{PhysRevLett.112.103604}, measurements of material properties~\cite{kaiser_quantum_2018}, as well as medical and biological sensing~\cite{crespi_measuring_2012}. The common concept in these applications lies in determining relative time delays accurately, as required for precise $\Delta n$ measurements. To date, the key ingredients for obtaining the highest precision time-delay measurement using the HOM effect are : i) the common-path geometry that significantly helps the stability of the interferometer while at the same time limiting the application only to birefringent samples~\cite{PhysRevA.62.063808,dauler1999}, and ii) the use of very short samples that do not exceed the coherence length of the single photons (less than 100~$\mu$m)~\cite{attosecond,biphoton-beat-note}.\\
 Here, we propose practical $\Delta n$ measurements based on QOCT in a dual-arm configuration with a 50 cm-long sample. The method aims at measuring the time-delay between two optical paths, each associated with a given material to be characterized. It is worth noting that HOM interferometry is immune to relative phase fluctuations between the two arms avoiding complex and expensive stabilization systems, as typically experienced in classical interferometry. Moreover, stringent conditions (identical length and temperature) are set for the two materials thanks to a special two-core rod-type fiber packaging.
 
\section{Results}
\begin{figure}[htp]
\centering
\includegraphics[width = 0.48\textwidth]{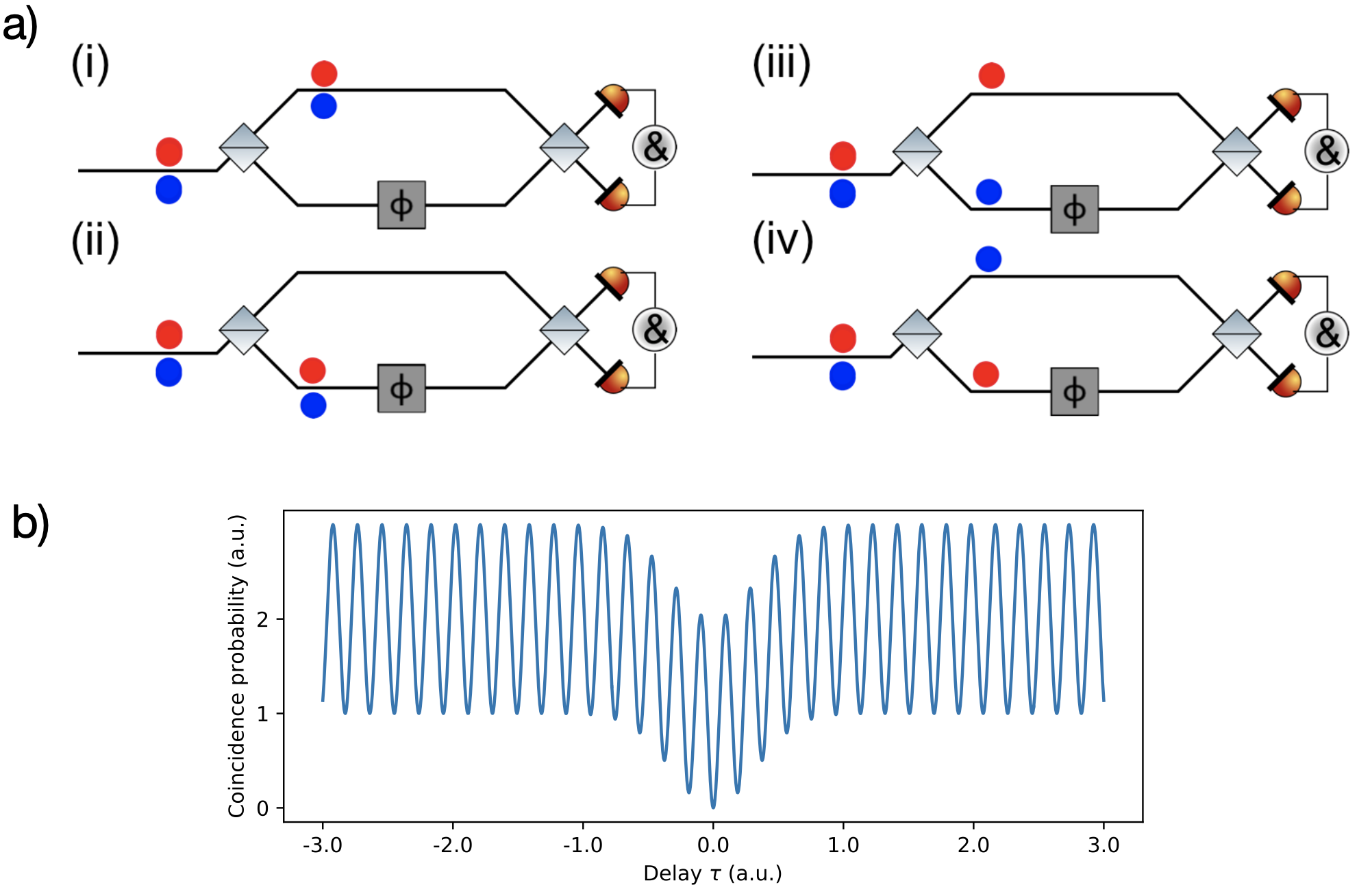}
\caption{a) Different case scenarios in two-photon interferometry using a MZ-type device. The pair can travel along four different paths. Interference in the coincidence counts can occur provided two paths are indistinguishable. Note that the different colors for the two photons are for representation purpose only. Ideally the paired photons are indistinguishable. b) Interference pattern at the output of a balanced interferometer}
\label{fig_differnt_paths}
\end{figure}

\textbf{Theory of two-photon interference.} A brief overview of the evolution of a two-photon state through a Mach-Zehnder interferometer (MZI)~\cite{lopez-mago_coherence_2012} that lies at the heart of our measurement method, is depicted in Fig.~\ref{fig_differnt_paths}.a.


Energy-time entangled photon-pairs generated by spontaneous parametric down conversion (SPDC) out of a second-order nonlinear crystal are here considered. Such 3-waves mixing process is ruled by energy and momentum conservation, written as $\omega_{\rm{p}}=\omega_{\rm{i}}+\omega_{\rm{s}}$ and $\vec{k_{\rm{p}}}= \vec{k_{\rm{i}}} + \vec{k_{\rm{s}}}$, respectively, where p, i, s refer to the pump, idler and signal photon, respectively. Their state can be written as
\begin{equation}
\ket{\Psi_{in}}=\int \text{d}\omega_{\rm{i}} \text{d} \omega_{\rm{s}} G(\omega_{\rm{i}}, \omega_{\rm{s}}) a^\dagger_{\omega_{\rm{s}}} a^\dagger_{\omega_{\rm{i}}} \ket{0},
\end{equation}
where $a^\dagger_{\omega_{\rm{s}}}$ ($a^\dagger_{\omega_{\rm{i}}}$) is the creation operator of a photon in input mode $a$ at frequency $\omega_{\rm{i}}$ ($\omega_{\rm{s}}$). $G(\omega_{\rm{i}}, \omega_{\rm{s}})$ and $|G(\omega_{\rm{i}}, \omega_{\rm{s}})|^2$ are the joint-spectral amplitude and density, respectively. The latter corresponds to the probability of detecting one photon at frequency $\omega_{\rm{i}}$ and the other one at $\omega_{\rm{s}}$. One should note that $\int \text{d}\omega_{\rm{i}} \text{d} \omega_{\rm{s}} |G(\omega_{\rm{i}}, \omega_{\rm{s}})|^{2}=1$. The joint spectrum amplitude of the generated biphotons in CW regime using a laser at frequency $\omega_p$ is given by $G(\omega_{\rm{i}}, \omega_{\rm{s}})=g(\omega_{\rm{i}}) g(\omega_{\rm{s}}) \delta(\omega_{\rm{p}} - \omega_{\rm{i}} - \omega_{\rm{s}})$. Both photons pass through the same bandpass filter which is generally added to clean the photons spectrum from spurious frequency components (see \cite{tanzilli_ppln_2002} for more details on experimental SPDC sources). The exact form of $g(\omega)$ depends on both the phase-matching condition and the transmission profile of the filter. In the case of a Gaussian shaped filter, $g(\omega) = (2\pi \sigma)^{-1/4} \cdot \text{e}^{\frac{(\omega-\omega_p/2)^2}{4 \sigma2}}$, with bandwidth $\sigma$ centered around $\omega_{\rm{p}}/2$. 

Sending such a state to an MZI (see ~Fig.\ref{fig_differnt_paths}), the probability of detecting two-photon coincidences between the two output ports of the device, as a function of the ajustable delay $\tau$ between its two arms reads~\cite{lopez-mago_coherence_2012}:
\begin{equation}
P_c(\tau)=P_{\rm{c}}(0)(2 - \cos \omega_{\rm{p}} \tau - \alpha \text{e}^{\tau^{2} \sigma^{2}}),
\label{eqn_P_c_1}
\end{equation}

\hspace{-0.5 cm} where $\alpha$ and $P_{\rm{c}}(0)$ represent the HOM-dip visibility and the average probability of registering two-photon events, respectively.

Interference in coincidence counting occurs between probability amplitudes of indistinguishable paths. In Eq. (\ref{eqn_P_c_1}) one can identify three terms. The first term is a constant, stemming from all of the possible distinguishable paths. The second term is related to two-photon contribution in superposition of travelling along the same path (case (i) and (ii) in Fig. \ref{fig_differnt_paths}.a, resulting in a Franson-type oscillation due to the interference of the so-called N00N-state with N=2~\cite{franson_bell_1989}. Such a two-photon state enhances the phase sensitivity by a factor $N$(2), Heisenberg-limited in precision \cite{Dowling2008}. This results in an interference pattern oscillating at the pump frequency $\omega_p$, instead of the central frequency of the single photons as it would be the case with classical light. The third term comes from two photons experiencing different arms (case (iii) and (iv) in Fig. \ref{fig_differnt_paths}.a and the interference of those two identical single modes (over all observables) at the second beamsplitter of the interferometer. This is equivalent to the HOM effect and results in a dip in the coincidence counts, which shape and width depends on the spectral amplitude $g$ of the photons. As a result, the figure of merit associated with Eq.~\ref{eqn_P_c_1} is a superimposition of a HOM-dip over Franson-type interferograms, as shown by Fig. \ref{fig_differnt_paths}.b  It must be emphasized that the precision on the path difference measurement is directly related to the spectral bandwidth of the photons. The broader they are, the narrower is the HOM-dip, and therefore better is the precision.\\
 In classical OCT, the intensity $I(\tau)$ at one of the output ports as a function of the path difference $\tau$ reads:
\begin{equation}
I(\tau)= I_0[1-V \cdot \cos(\omega_{\rm{c}} \tau) \cdot f(\tau)],
\label{eqn_standard}
\end{equation}

\noindent where $I_0$ is the average intensity, $V$ the experimental visibility, $\omega_{\rm{c}}$ the central frequency on the interferogram, and $f(\tau)$ an envelope function which depends on the spectral width and shape. Without dispersion,  $f(\tau)$ reaches its maximum value of 1 for $\tau=0$ since all frequency components within the SPDC spectrum arrive simultaneously at the second beamsplitter and interfere. With dispersion, the different frequencies arrive at different times, resulting in a reduced visibility and in a larger envelope function $f(\tau)$. Dispersive effects therefore reduce the achievable precision in determining optical path differences equality in OCT. On the other hand, HOM-interferometry is insensitive to even order dispersion, including the dominant term of chromatic dispersion \cite{mozzotta2016,okano2013,okano2015, abouraddy_quantum-optical_2002}. The related experimental visibility $\alpha$ (Eq. \ref{eqn_P_c_1}) only depends on the indistinguishability of the two photons (in terms of time, polarization, and spatial mode). This also induces a robust immunity against propagation losses added by the sample under test, which is not the case in standard OCT. Hence, the quantum approach is fully independent on the sample under test characteristics (chromatic dispersion, losses) and thus permits reliable, practical, and high-precision measurements with the perspective of addressing actual quantum metrology scenarios.\\

\begin{figure*}[htp]
\centering
\includegraphics[scale=0.4]{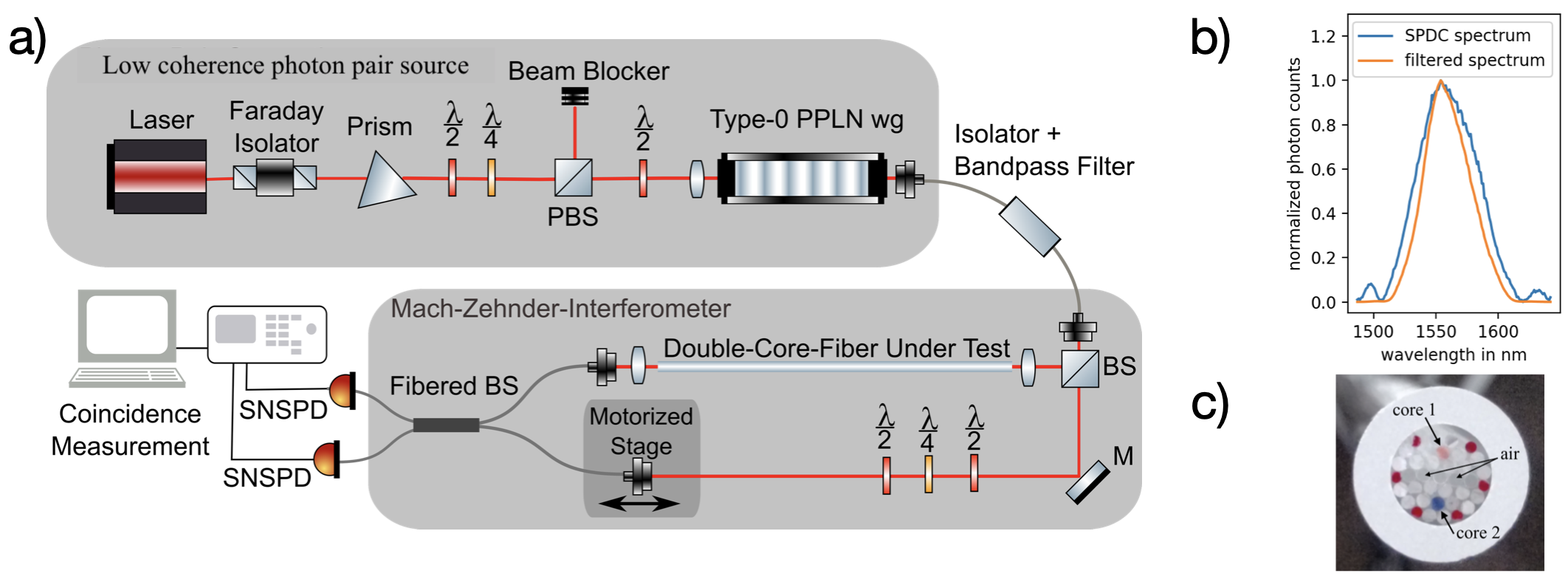}
\caption{a) Experimental setup. A periodically poled lithium niobate waveguide (PPLN-wg) is pumped at 780~nm (CW laser) to generate entangled photon pairs. Those are spectrally bandpass filtered (BPF) and sent to a home-made Mach-Zehnder interferometer. One arm is adjustable and the other contains the two-core rod-type fiber sample. The two output modes are directed to two superconducting nanowire single-photon detectors (SNSPD) that are connected to a time-to-digital converter (TDC) to record the coincidence counts. b) Measured SPDC spectrum with and without the $\SI{90}{nm}$ passband filter. Both curves are normalized with respect to their maximum. We find a FWHM of $\SI{44}{nm}$. c) Cross-section of the special two-core fiber. There is an low refractive index barrier preventing between the two cores to avoid evanescent coupling between them. Each core has a diameter of $\sim \SI{10}{\mu m}$ and they are separated by $\sim \SI{30}{\mu m}$. \textit{Core 1} and \textit{Core 2} refers to the material constituting the core and the cladding of an VLMA fiber, respectively.}
\label{fig_setup}
\end{figure*}

\begin{figure}[htp]
\centering
\includegraphics[width = 0.48\textwidth]{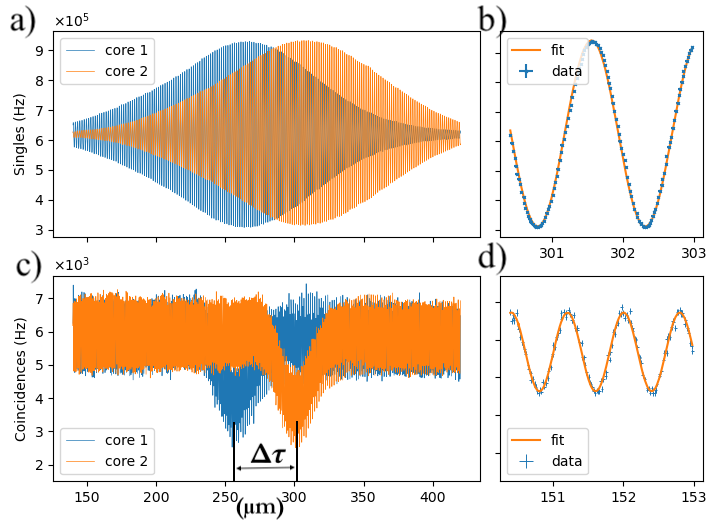}
\caption{a) Measured photon counts at one output port as a function of the delay $\tau$. b) Zoom of the central region allowing to resolve the phase fringes at $\lambda=\SI{1560}{nm}$. The fit permits to infer a visibility of 0.5. c) Measured coincidence counts between the two output ports as a function of the delay $\tau$. d) Zoom of the central region allowing to resolve the Franson-type oscillation at $\lambda=\SI{780}{nm}$. The fit permits to infer a visibility of 0.74. All data are measured with an acquisition-time of 1~second per point.}
\label{fig_interferogram}
\end{figure}

\textbf{Refractive index differences measurement}. We aim at measuring $\Delta n$ between two materials constituting the core and the cladding of a VLMA fiber. These two materials are embedded into a special two-core (one material for each core) rod-type fiber. The measured interferogram, obtained thanks the experimental setup represented in Fig. \ref{fig_setup}, for one core is shown in Fig. \ref{fig_interferogram}, for both OCT and QOCT methods. Similar patterns are obtained for the second core with an offset $\Delta \tau$. A detailed description of both the experimental setup and the methodology is provided in the “Methods”section.

\begin{figure}[htp]
\centering
\includegraphics[width = 0.48\textwidth]{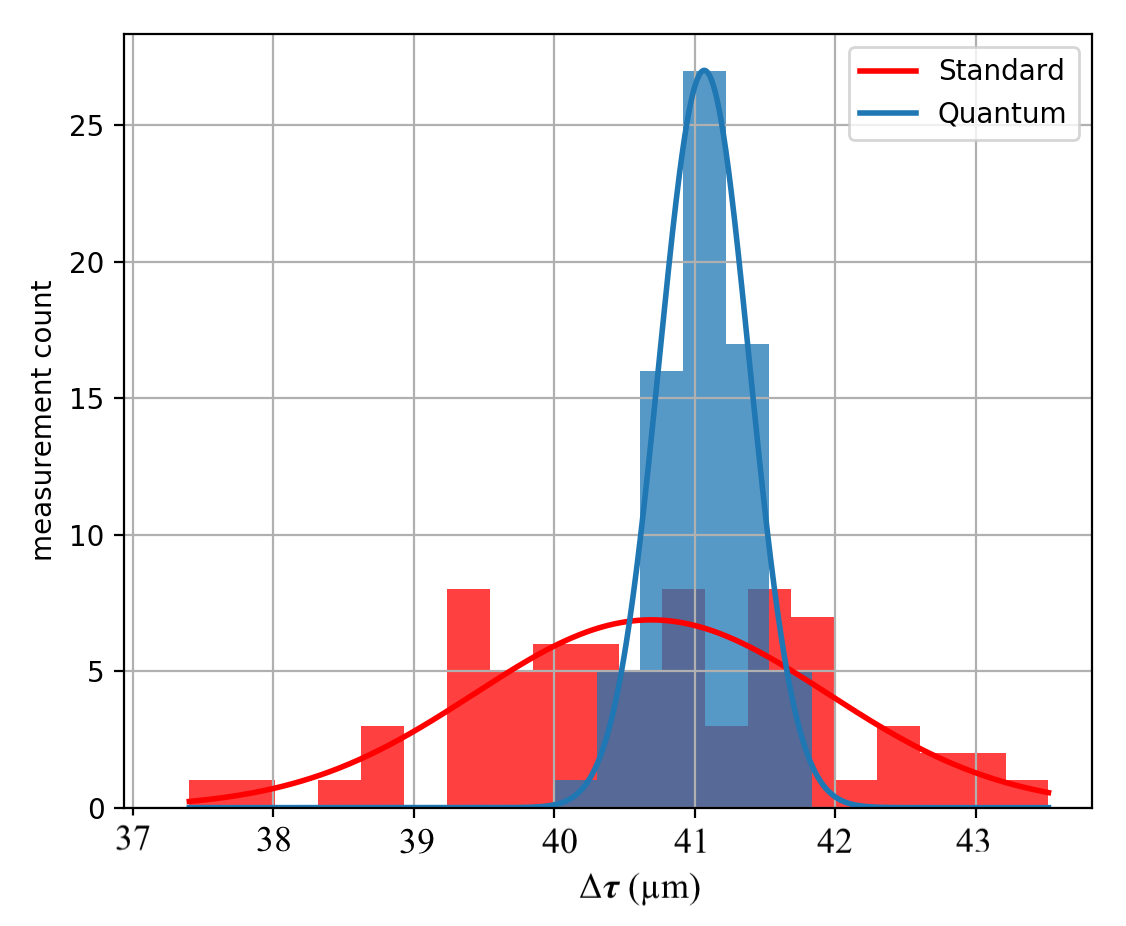}
\caption{Histogram of inferred index difference after 70 repetitions with the same two-core fiber for both standard (red) and quantum-enhanced (blue) measurements. Fits to the data assumed a normal distribution.}
\label{fig_histo}
\end{figure}

Prior to the estimation of the precision with both methods, we evaluate the expected enhancement with HOM-interferometry in comparison with the OCT. For the OCT-interferogram measurement, we fit the oscillation of the single counts shown in Fig.~\ref{fig_interferogram}.a according to Eq.~\ref{eqn_standard}. A visibility and a FWHM equal to $V_{OCT}=50 \%$ and $\SI{134}{\mu m}$ are respectively inferred. A zoom showing both the experimental and fitted OCT-interferograms is shown in Fig.~\ref{fig_interferogram}.b.  This reduced visibility mainly comes from the propagation losses of the two-core sample.\\
Similarly, we fit the experimental HOM-dip shown in Fig.~\ref{fig_interferogram}.c using Eq.~\ref{eqn_P_c_1}. The corresponding raw visibility and FWHM of the HOM-dip are deduced from the fitting curve (Fig.~\ref{fig_interferogram}.d) and are equal to $V_{QOCT}=74 \%$ and $\SI{25.8}{\mu m}$, respectively. For the $\SI{44}{nm}$ wide spectrum of the entangled-photon pairs one would expect a FWHM of $\SI{21.7}{\mu m}$. This enlargement of 19\% comes from the third-order dispersion, resulting in a slightly asymmetry but keeping its integral constant \cite{mozzotta2016,okano2013}. This broadening hence limits the expected $V_{QOCT}$ by a similar amount, \textit{i.e} to 74\%. As a consequence, the visibility can no longer be considered as a criterion for the indistinguishability between two photons in the presence of odd higher-order dispersion. In this case, we rather have to compare the theoretical integral of the HOM-dip, which is obtained by Fourier transform of the Gaussian bandpass filter, to that of the experimental HOM-dip. This ratio is equal to 94 \% and corresponds to the raw equivalent HOM-dip visibility subtracting third-order dispersion contribution. Furthermore, this non-unit ratio is explained by a non-perfect mode matching between the two input photons and by a slightly unbalanced beam-splitter.\\
The precision is mainly given by the coherence length of the source, that is inversely proportional to its spectral bandwidth. The HOM-dip width is 5-times shorter than the envelope function of the classical interferogram. The chromatic dispersion broadens the classical interferogram, while the HOM-dip stays essentially unaltered, since the visibility only depends on the indistinguishability of the two photons. The robust behaviour of the quantum approach lies at the heart of the enhancement attained through the use of entangled photons instead of classical light.\\ 
The achievable precision of $\Delta n$ measurements mainly depends on the width of the interferogram but also on its intensity fluctuations. There is a factor 100 between the coincidence and the single-photon counts. This comes from the overall losses of about 20 dB from the output of the dual-core fiber to the output of the interferometer. The major contributions come from  the coupling from free space to fibers and the injection into the two-core fiber. Since coincidence and single-photon counts follow a Poisson-statistic \cite{HAYAT1999275}, there are $\sim 10$ times more fluctuations due to the shot-noise in the quantum compared to the classical measurement. Consequently, one can expect a little bit less then a 5-fold enhancement in precision between the classical and quantum methods.\\
 All measurements were repeated 70 times to infer the statistical accuracy of both approaches. We switch every time between the two cores of the two-core fiber in order to keep the same environmental conditions (essentially the temperature) during the overall experiment. The results of the statistical data analysis are shown in Fig. \ref{fig_histo}. We obtain $\Delta \tau^{\text{OCT}}=\SI{40.7(12)}{\mu m}$, which outperforms any classical measurement by one order of magnitude~\cite{Tan:16,singh2006} and $\Delta \tau^{\text{QOCT}}=\SI{41.1(3)}{\mu m}$ corresponding to OCT and QOCT approaches, respectively. This corresponds to a $\Delta n$ precision equal to $\sigma_{\Delta n}^{\text{OCT}}=\num{24e-7}$ and $\sigma_{\Delta n}^{\text{QOCT}}=\num{6e-7}$. This precision, standing as the highest achieved in terms of $\Delta n$ fits well our expectation considering the interferogram width and fluctuations associated with the counts statistic. This enhancement results as a clear manifestation of the peculiar properties of energy-time entangled photon pairs, allowing for chromatic dispersion cancellation~\cite{abouraddy_quantum-optical_2002}. This work therefore demonstrates that such quantum advantages are of high interest for characterizing optical samples without having any prior knowledge on their properties. This becomes even more interesting when working with realistic or long samples.\\
The origin of the standard deviations in Fig.~\ref{fig_histo} arises from several reasons. Due to the switching-method between the two cores in order to measure index differences in the same conditions, mechanical drifts cause ineluctably small systematic errors in the optical length. In order to minimize these fluctuations, the positions of input- and output-lenses are fixed, hence keeping the same focus points throughout the overall duration of the experiment. We prefer moving the fiber on both extremities in order to align them within the focus point of the lenses. That way, the angle alignment error is minimized.
Furthermore, thermal fluctuations play an important role for all kinds of interferometric methods, especially when long samples are involved. Note that temperature variations of $\Delta T \sim\SI{0.1}{K}$ result in drifts on the order of one-phase fringe in the quantum measurement. Since we are working in laboratory conditions and both cores are contained within the same rod, we have verified that our system is more stable than 0.1~K within the recording time. The total duration for the overall measurement takes 8 hours corresponding to all the necessary data for the histogram in Fig. \ref{fig_histo}, which encompasses both the variance due to thermal and mechanical fluctuations, and also from the estimation method (see the appendix).\\
A further enhancement in precision is possible, when using for example a larger SPDC spectrum \cite{okano2015} or techniques exploiting a maximum-likehood estimator, while pre-tuning the interferometer to the position that contains the maximum information content \cite{attosecond}.
Those demanding methods require ultra-precise active thermal stabilization, that imposes further technical challenges. Our method stands as a trade-off between practicability and precision, that does not require complex implementation of active stabilization systems while still achieving high-precision and being user-friendly.

\section{Discussion}
In this paper we have implemented an experimental method based on two-photon interference, referred to QOCT, to measure index difference between two materials that are embedded within the same fiber. Using HOM-interferometry and large frequency-entangled photon pairs we achieved unprecedented precisions up to $\sigma^{\text{QOCT}}=\num{6e-7}$. We compared the QOCT and OCT approach. Even though we already achieved ultra precise results using standard approach we still found a 4-fold enhancement in precision for the QOCT measurement due to both even term dispersion insensitivity  and robustness to the loss in HOM-interferometry. Our precise results will find use in various fields, notably for special large-mode-area fibers that are crucial for the development of powerful fiber lasers in the future. 

\section{Methods}
\textbf{Experimental setup} The experimental setup is shown in Fig.~\ref{fig_setup}.a. A continuous-wave laser operating at $\SI{780}{nm}$ pumps a type-0 periodically poled lithium niobate waveguide (PPLN-wg) that produces, via SPDC, degenerated, broadband, energy-time entangled photon pairs. Fig. \ref{fig_setup}.b shows the spectral density (corresponding to $|g(\omega)|^2$) at the output of the PPLN-wg. The side peaks are discarded thanks a $\SI{90}{nm}$ passband filter, centered at $\SI{1560}{nm}$. The filtered spectrum can be fitted by a Gaussian function, with a full width at half maximum (FWHM) of $\SI{44}{nm}$.

The choice for a type-0 phase-matched source is motivated by its natural broadband SPDC-spectrum, \textit{i.e} low-coherent photons. 
The generated photons are then sent to a Mach-Zehnder interferometer. The reference is adjustable via a nano-positioning stage, ranging from $\num{0}$ to $\SI{500}{\mu m}$  and having a precision of $\SI{20}{nm}$. The other arm contains the sample, being a special two-core fiber. Piezo-actuators allow fast switching from one core to the other in the tranverse plane to the waveguide axis. The two cores are made of different materials corresponding to those constituting the core and the cladding of the VLMA (see Fig. \ref{fig_setup}.c), leading two optical paths. The cores are separated by an air-gap to prevent any coupling between them. Note that the fiber is actually embedded into a solid rod to avoid systematic errors arising from the fiber-curvature and/or polarization drifts. A single-mode fiber beamsplitter recombines the signal from both arms of the interferometer to guarantee the projection onto identical spatial modes. Moreover, a polarization controller ($\lambda/2$-, $\lambda/4$-, $\lambda/2$-waveplate) in the free-space arm is added in order to ensure the indistinguishability of the polarization modes.\\
The experimental method consists in coupling quantum light in one of the cores, performing a fast scan ($\sim$ min) of the HOM-dip, and then repeat this procedure after switching to the other core. Since each core is made of a different material, the centers of the two interferograms have an offset $\Delta \tau$ that corresponds exactly to the optical-path difference between the two cores. Knowing the exact physical length $L=\SI{50 \pm 0.01}{cm}$ of the sample, one can deduce the index difference between the two cores, given by $\Delta n = \frac{\Delta \tau} {L}$. Advantageously, the precision measurement is shifted from $\Delta n$ to our ability of determining an optical-path difference in th etime domain $\Delta \tau$ with a high precision. The value of $\Delta n$ is related to the group-index difference including both material and waveguide contributions. The latter contribution can easily be evaluated and then removed thanks to standard simulations in order to infer the index difference between the two materials~\cite{1464-4258-8-11-001}.\\
\textbf{Post-data analysis} In the perspective of determining the optical-path difference $\Delta \tau$ between the two cores, a Fourier-transform based estimator~\cite{Diddams:96} is implemented based on its property under translation
$\mathcal{FT}_x [f(x+ t_0)](\omega) = \mathcal{TF}_x[ f(x)](\omega) \cdot \text{e}^{\text{i} \omega t_0}$. As described in details in the supplementary information section, the offset $\Delta \tau$, that corresponds to the delay between the two HOM-dips, is deduced from a linear fit of the spectral phase of the interferogram at low frequencies (corresponding to the HOM-dip). In order to fairly compare the QOCT and OCT methods, the spectral bandwidth of both sources has to be identical. We simultaneously exploit coincidence counts and  single-photon at one of the output ports for the QOCT and OCT approaches, respectively. Furthermore, as for the quantum approach, we apply a similar Fourier-transform based estimator, now fitting the phase around the central frequency of single photons. A detailed description of both quantum and classical estimation methods can be found in the appendix.\\
In order to fairly estimate the precision, we switch 70 times between the two cores of the same sample, estimating each time $\Delta \tau$ via quantum and classical methods to infer the statistical accuracy of both approaches. 

\section*{Acknowledgment}
This work has been conducted within the framework of the project OPTIMAL granted by the European Union by means of the Fond Européen de développement regional (FEDER). The authors also acknowledge financial support from the Agence Nationale de la Recherche (ANR) through the projects METROPOLIS, the CNRS through its program “Mission interdisciplinairité” under project labeled FINDER, and the French government through its Investments for the Future programme under the Université Côte d'Azur UCA-JEDI  project  (Quantum@UCA)  managed  by  the  ANR (ANR-15-IDEX-01).\\
The authors also acknowledge technical support from IDQ. We also thank Elie Gouzien and Yann Bouret for assistance with data post-processing.

\section*{Author information}

M.R., F.M., D.A. and R.C. performed the experiments. R.D., B.L., P.R. and R.J. were in charge of designing and fabricating the two-core fiber. M.R., F.K. L.L. and S.T. designed the experiment. M.R., L.L. and S.T wrote the paper with inputs from P.R. and R.J.

\section*{Competing interests}
The authors declare that there are no competing interests.

\section*{Data Availability}
Data are available from the authors on reasonable request.

\footnotesize
\bibliographystyle{ieeetr}

\clearpage
\onecolumngrid
\normalsize

\appendix
\addcontentsline{toc}{section}{Supplementary information}

\appendix

\vspace{50px}
\section*{Supplementary information}
\subsection{ Fourier-based estimators for the optical path difference between the two cores}

Since we measure each interferogram as a function of the motor-position $p=\tau+p_0$ and not directly as a function of the optical path length difference $\tau$ between the two arms of the interferometer, we have to find $p_0$, the motor-position that corresponds to the optical-path equality with the highest precision possible. Therefore we use an estimator that is taking advantage of the property of the Fourier transform under translation. Let's suppose a function $u(\tau)$ and its Fourier transform with respect to $\tau$ is $\mathcal{FT}_\tau [u(\tau)]=\hat{u}(f)$, a function of the frequency $f$.
The Fourier transform of the translated (time-shifted) function $v(\tau)=u(\tau-\tau_0)$ can then be written as $\hat{v}(f)=\hat{u}(f) \cdot \text{e}^{- 2 \pi \text{i} f \tau_0}$ where an extra phase term is appearing that is evolving linear with the frequency $f$ with slope of $-2\pi\tau_0$. For an even function $u(\tau)$ , $\hat{u}(f)$ is a real function and the only phase term of $ \hat{v}(f)=\hat{u}(f) \cdot \text{e}^{- 2 \pi \text{i} f \tau_0}$ is $-2\pi f \tau_0$. Linear-fitting the phase of $\hat{v}(f)$, one thus can easily deduce $\tau_0$.

Due to small third-order (and higher-odd order) dispersive effects, the interferograms (for quantum and classical measurement) are not perfectly symmetric \cite{mozzotta2016,okano2013,Diddams:96}. The Fourier-transform of an unsymmetrical function $u(\tau)$ is complex $\hat{u}(f)= \left| \hat{u}(f) \right| \cdot \text{e}^{\text{i}\psi(f)}$, with a phase term $\psi(f)$ that is varying with the frequency $f$. The phase of the Fourier transform of the translated function $\hat{v}(f)=\left|\hat{u}(f)\right| \cdot \text{e}^{\text{i}\psi(f)} \cdot \text{e}^{- 2 \pi \text{i} f \tau_0}$ is thus a superposition of $\psi(f)$ and the linear term $- 2 \pi f \tau_0$. A simple linear-fit cannot determine $\tau_0$.

Fortunately we are mainly interested in the difference $\Delta \tau$ of the optical path between the two cores and not the precise motor-position, that corresponds to the path equality for each core. Calculating the Fourier transform of the interferogram for each core, its phase for core 1 (core 2) takes the form $\Psi^{(1)}=\psi^{(1)}_{\text{asym}}(f) - 2 \pi f \tau^1_0$ ($\Psi^{(2)}=\psi^{(2)}_{\text{asym}}(f) - 2 \pi f \tau^2_0$), where $\psi^{(1)}_{\text{asym}}(f)$ ( $\psi^{(2)}_{\text{asym}}(f)$) stands as the phase terms resulting form the unsymmetrical interferogram for core 1 (core 2). We suppose that the interferograms for the two cores are only translated in respect to each other, but the form of the dip/ envelope is the same, resulting in $\psi^{(1)}_{\text{asym}}(f))=\psi^{(2)}_{\text{asym}}(f)=\psi_{\text{asym}}(f))$. This is a valid approximation since we use the same spectra. Calculating the difference $\Psi^{(2)}-\Psi^{(1)}=\psi_{\text{asym}}(f) - 2 \pi f \tau^2_0 - (\psi_{\text{asym}}(f) - 2 \pi f \tau^1_0 )=2 \pi f (\tau_1 - \tau_2)$, the phase terms resulting form the unsymmetrical interferograms cancels out and with a linear-fit we have directly access to $|\tau_1 - \tau_2|=\Delta \tau$.

\begin{figure}[tp]

\centering

\includegraphics[width=\textwidth]{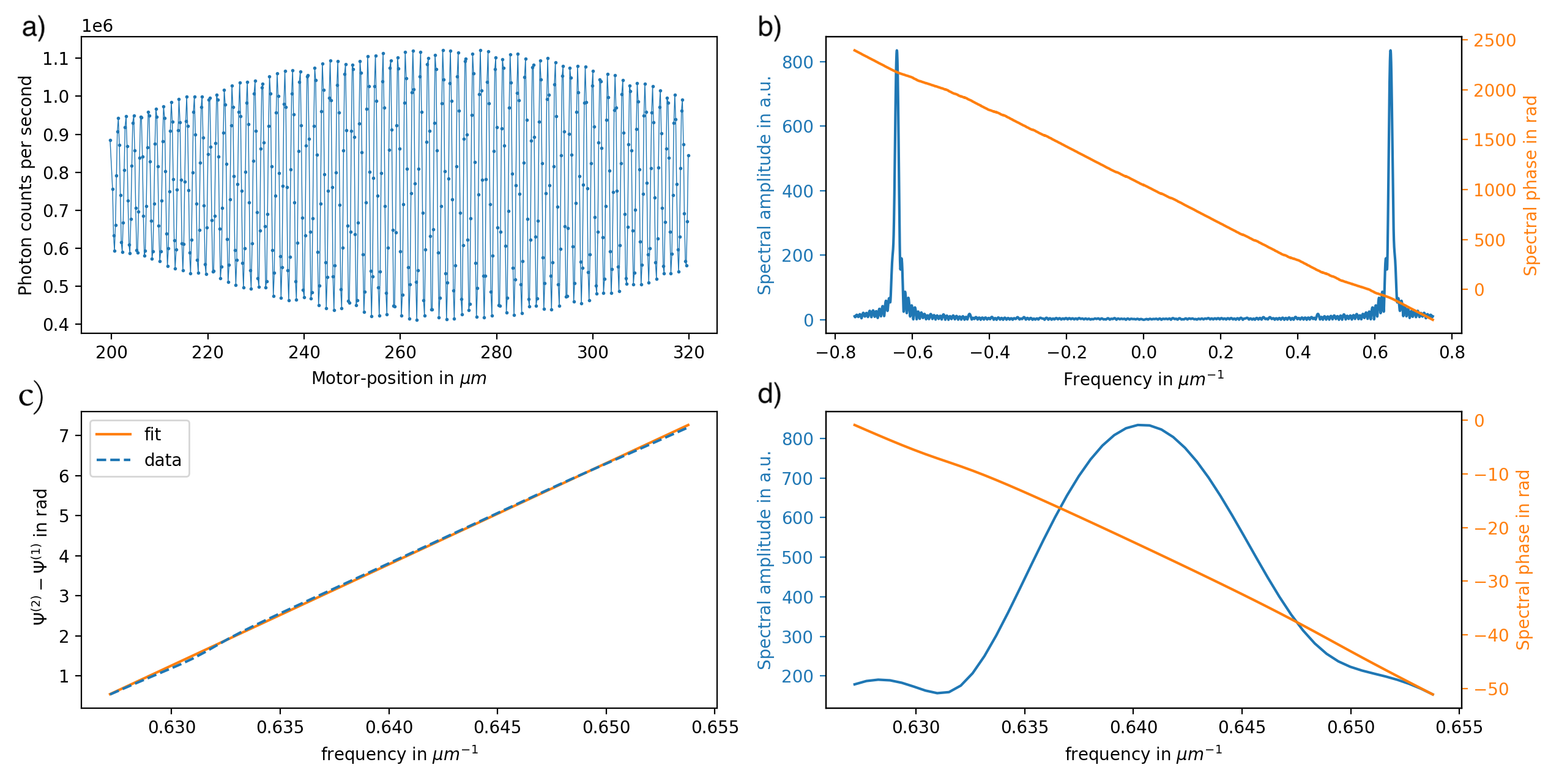}
\caption{a) Measured classical interferogram with a step-size of $\SI{0.24}{\mu m}$ and an integration time of $\SI{0.5}{s}$. b) amplitude and phase of the Fourier transform of the classical interferogram. c) amplitude and phase of the Fourier transform, zoomed-in around side peak at $\frac{1}{\SI{1560}{nm}}$. d) Differential phase $\Psi^{(2)}-\Psi^{(1)}$ of the Fourier transform for two different measurements (each in a different core) with a linear fit to find $\Delta \tau=\SI{40.9 \pm 0.4}{\mu m}$.} 
\label{fig_estimator_cl}
\end{figure}

\begin{figure}[tp]
\centering

\includegraphics[width=\textwidth]{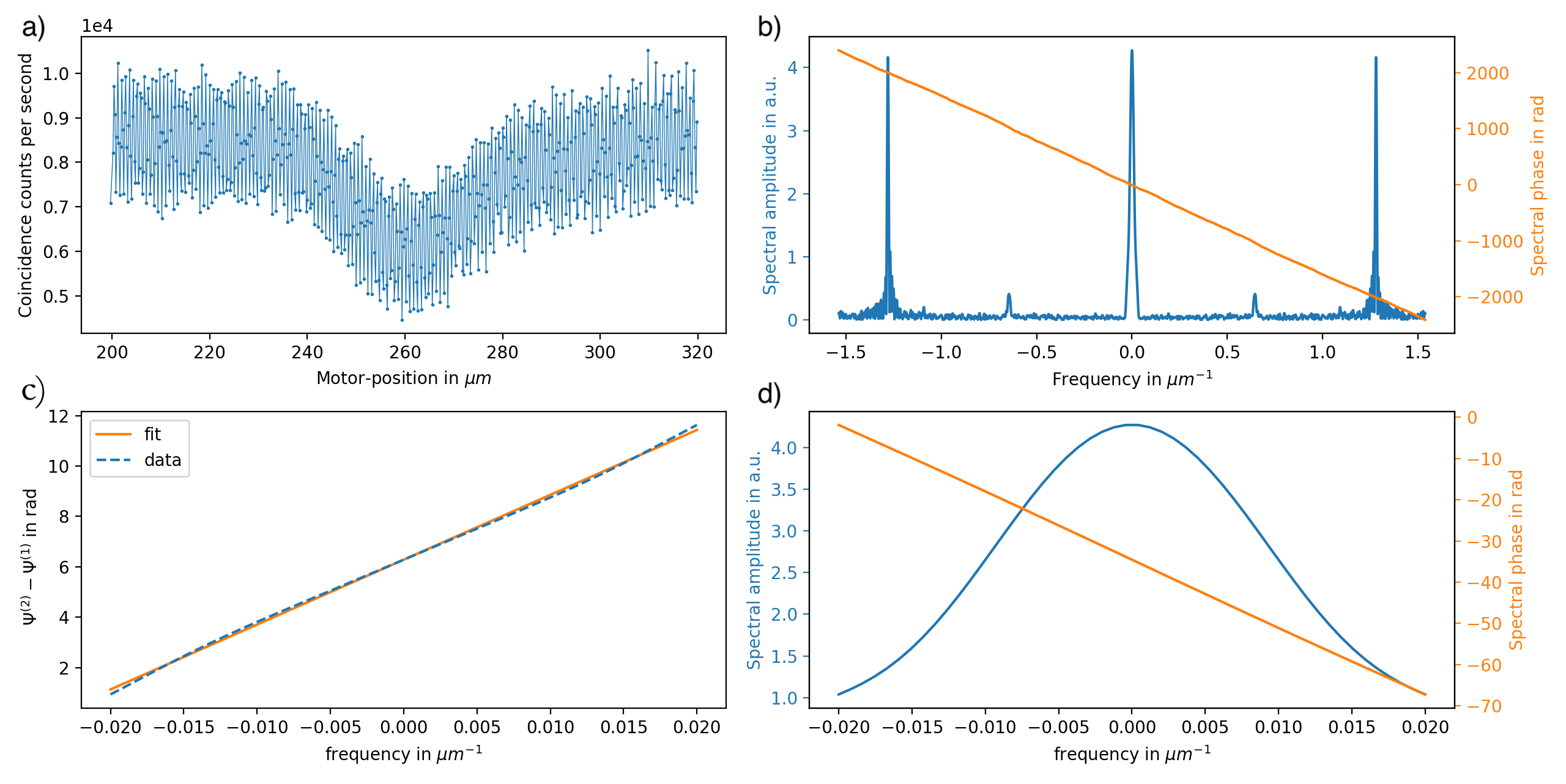}
\caption{a) Measured quantum interferogram with a step-size of $\SI{0.24}{\mu m}$ and an integration time of $\SI{0.5}{s}$. b) amplitude and phase of the Fourier transform of the quantum interferogram. c) amplitude and phase of the Fourier transform, zoomed-in around central peak, corresponding to the HOM-dip. d) Differential phase $\Psi^{(2)}-\Psi^{(1)}$ of the Fourier transform for two different measurements (each in a different core) with a linear fit to find $\Delta \tau=\SI{41.3 \pm 0.3}{\mu m}$.} 
\label{fig_estimator_qu}
\end{figure}

Since the interferogram is a discrete set of data-points, that are non uniformly distributed due to the variance of $\sim \SI{20}{nm}$ in the motor-position, we mathematically calculate its Fourier-transform with a nonuniform discrete Fourier transform (NDFT) algorithm (Fig.~\ref{fig_estimator_qu} and \ref{fig_estimator_cl}).

We perform prior simulations of the expected interferogram according to (\ref{eqn_P_c_1}), supposing Gaussian distributed fluctuations within the motor-position and a Poisson distribution for the photon counts/coincidences in order to find the optimal stepsize and total range of the HOM-dip scan. This leads to a trade-off between the total number of points and the resulting precision of the estimator, since the faster we can perform the scan the less we are sensible to temperature fluctuations during consecutive measurements. We can reduce the total number of points to 500, before the precision degrades. This gives us a total range of $\SI{120}{\mu m}$, centered around the prior estimated optical path equality, with a step-size of $\SI{0.24}{\mu m}$. With an integration time of $\SI{0.5}{s}$ per point this results in a measurement time of $\sim 5$ minutes per core. 

\subsection{Estimator for the HOM and classical white-light interferometry}
A measured interferogram and its calculated Fourier transform can be seen in Fig.~\ref{fig_estimator_qu} and \ref{fig_estimator_cl}, corresponding to the quantum and classical regime respectively. Before calculating the Fourier transform, we subtract the average outside the dip~\cite{bracewell_fourier_2000}. In the quantum approach, we therefore filter the Fourier transform around the low-frequencies that correspond to the HOM-dip.\\
On the other hand, in the classical regime, the envelop function is modulated by the single-photon frequency, we thus filter the Fourier transform around the side peak at $\frac{1}{\SI{1560}{nm}}$ in order to estimate $\Delta \tau$. Since the interferogram is a real function, its Fourier transform is symmetric, we therefore do not gain any more information by using the two side-peaks at $\frac{1}{\SI{1560}{nm}}$ and $-\frac{1}{\SI{1560}{nm}}$.\\
In both cases, we find a good agreement between the differential phase $\Psi^{(2)}-\Psi^{(1)}$ and the linear fit (Fig.~\ref{fig_estimator_qu} (c) and \ref{fig_estimator_cl} (c)), which is as well promising for the precision of the classical estimator.\\

\end{document}